\documentclass[10pt,twocolumn]{article}
\usepackage[margin=0.72in]{geometry}
\usepackage[T1]{fontenc}
\usepackage[utf8]{inputenc}
\usepackage{lmodern}
\usepackage{amsmath,amssymb}
\usepackage{graphicx}
\usepackage{booktabs}
\usepackage{enumitem}
\usepackage{abstract}
\usepackage{caption}
\usepackage{hyperref}
\usepackage{xcolor}
\usepackage{microtype}

\hypersetup{colorlinks=true, linkcolor=black, urlcolor=blue, citecolor=black}
\newcommand{\um}{\textmu{}m}

\title{The Need for Neural ISP in the Small-Pixel Era: How Shrinking Pixels Push Optics to the Limit and Neural Restoration Pushes Back}
\author{
Jingxi Li \quad
Neerja Aggarwal \quad
Laurent Gudemann \quad
Shivansh Rao \quad
Vishal Vinod \\
Tom E. Bishop \quad
Ziv Attar\\[2pt]
Glass Imaging Inc.
}
\date{}

\begin{document}
\maketitle

\begin{abstract}
Smartphone telephoto cameras are approaching a ``telephoto physics wall'': as pixel pitches shrink toward sub-0.5~\um, the optics remain limited by geometric aberrations, leading to diminishing returns on resolution. Traditional Image Signal Processors (ISPs) cannot eliminate these aberrations, because they operate through local, stage-wise processing with no explicit model of the underlying point spread function (PSF). We demonstrate how a learning-based Neural ISP for image restoration, trained on the underlying degradations, inverts what stage-wise pipelines cannot, turning small-pixel designs into a net advantage.

We investigate this through a controlled simulation of a representative telephoto module, evaluating five configurations (0.35--0.75~\um{} pixel pitch).  The aperture is scaled proportionally to keep per-pixel SNR and diffraction spot size fixed, thereby isolating geometric aberration and spatial sampling. While the traditional ISP improves only modestly with smaller pixels, the Neural ISP scales substantially: at 0.35~\um{} it reaches 745 cycles/mm MTF50 (vertical), a 2.5--3$\times$ resolution improvement over the traditional ISP, and LPIPS improves significantly from 0.244 to 0.151 while traditional results stay comparatively flat. In a low-SNR extension (15~dB per-frame bursts at 0.35~\um), a multi-frame Neural ISP recovers performance close to the bright-light single-frame baseline, whereas a multi-frame traditional ISP shows no meaningful improvement---indicating that traditional pipelines at small pixels are bottlenecked by uncorrected PSF blur rather than by noise. These results point to a design philosophy in which Neural ISPs enable high-resolution telephoto modules by correcting residual optical aberrations rather than requiring increasingly complex optics.

\end{abstract}

\section{The Telephoto Physics Wall}

\begin{figure*}[t]
    \centering
    \includegraphics[width=0.95\textwidth]{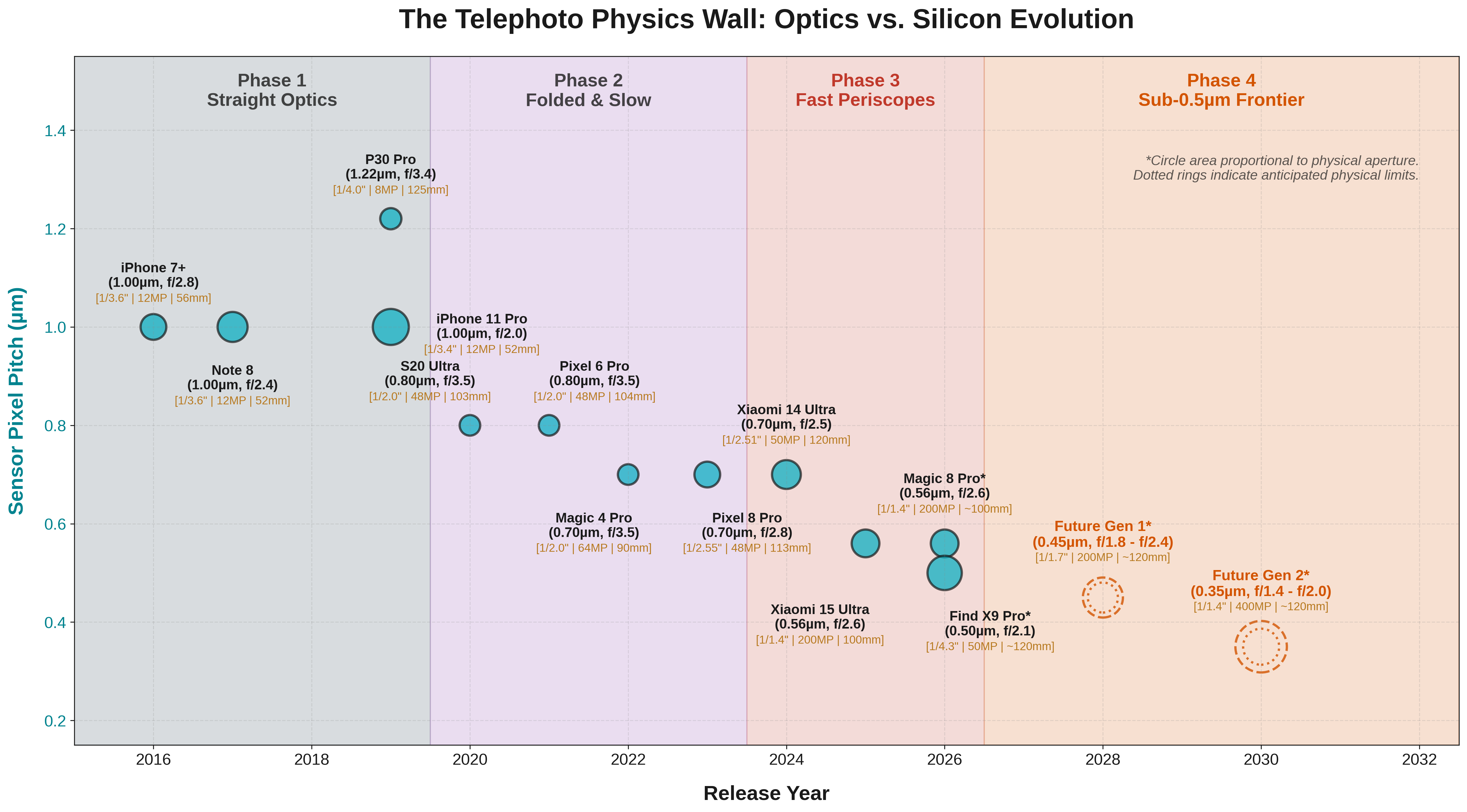}
    \caption{Telephoto sensors over the years show a decrease in sensor pixel pitch, and following the shift to folded designs, corresponding decreasing f/\# to pack more resolving power. Data source: GSMArena.}
    \label{fig:fig1}
\end{figure*}

Smartphone telephoto cameras have undergone a dramatic transformation over the past decade (see Figure~\ref{fig:fig1}). Pixel pitch has shrunk from 1.0~\um{} in the iPhone 7+ era down to approximately 0.5~\um{} in current 200MP periscope modules, with sub-0.5~\um{} designs already on the horizon. Each generation packs more resolving power into the same physical sensor area, continuing a broader trend in computational photography in which optical design and image reconstruction are increasingly co-designed~\cite{heide2014flexisp,sitzmann2018endtoend,tseng2021differentiable}.

As pixels shrink, the optics must keep pace. The lens must work harder to produce a smaller point spread function (PSF), i.e., the image of a single point source. To prevent diffraction blur from dominating at smaller pixel sizes, camera designers lower the F-number, widening the aperture to keep the diffraction spot small relative to the pixel pitch. This yields faster lenses (i.e., wider apertures allowing higher light collection and typically faster shutter speeds). The recent Oppo Find X9 Pro (0.5~\um, F/2.1) employs a significantly faster lens than the Honor Magic8 Pro (0.56~\um, F/2.6), and this trend is expected to continue.

Opening the aperture introduces a trade-off, however. A wider aperture captures rays from steeper angles at the edge of the pupil. These marginal rays contribute disproportionately to geometric aberrations---spherical aberration, coma, astigmatism, and higher-order terms that even the best lens designs struggle to eliminate. Thus while the diffraction spot size is reduced, aberration-induced blur increases with wider apertures. The result is what we term the ``telephoto physics wall'': the point at which silicon scaling outpaces what conventional optics and traditional ISPs can cleanly deliver, and further resolution gains are limited.

\begin{figure}[t]
    \centering
    \includegraphics[width=\columnwidth]{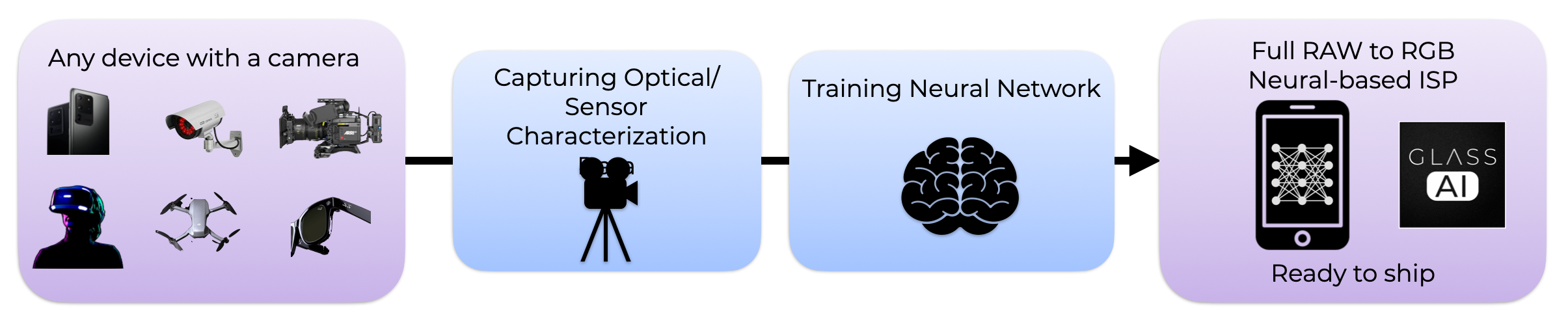}
    \caption{Glass Imaging's typical neural ISP training pipeline.}
    \label{fig:fig2}
\end{figure}

Over the past few years, Glass Imaging has developed a neural-based ISP, called Glass AI, designed to overcome this limitation. As shown in Figure~\ref{fig:fig2}, we train a camera-specific neural network to invert optical aberrations and sensor effects in real time on edge devices. We have evaluated this approach against state-of-the-art smartphone ISPs and demonstrated substantial improvements in image quality~\cite{glassblog_iphone16vs17}.

This raises a fundamental question: as pixels shrink, apertures widen, and aberrations grow, how effectively can a Neural ISP recover lost detail? Can Neural ISPs break the limits faced by traditional ISPs and demonstrate that pixel sizes below 0.5~\um{} and sensors beyond 200MP can still yield meaningful resolution benefits?

To answer this, we designed a controlled simulation study that isolates these effects. We simulated image formation through a set of representative smartphone lens modules with increasing aperture and corresponding CMOS image sensors with decreasing pixel size.

Our study demonstrates the necessity of using Neural ISPs to deblur severely aberrated PSFs as pixel size shrinks, enabling the next generation of high-resolution telephoto cameras. In contrast, traditional ISPs without lens-specific deconvolution exhibit the telephoto physics wall, with limited benefit from smaller pixels.

\paragraph{Contributions.} This work makes the following contributions:
\begin{itemize}[leftmargin=*]
    \item A controlled simulation protocol that scales aperture with pixel pitch ($\text{F/\#} \div \text{pitch} = \text{constant}$) to isolate geometric aberration and spatial sampling from diffraction and noise.
    \item A characterization of how traditional and Neural ISP restoration scale as pixel pitch shrinks from 0.75 to 0.35~\um, showing that the Neural ISP advantage grows with smaller pixels while the traditional ISP exhibits a ``physics wall.''
    \item A multi-frame low-light analysis showing that the two pipelines are bottlenecked by different effects: at small pixels the traditional pipeline is limited by uncorrected PSF blur rather than by noise.
    \item Design implications for next-generation telephoto modules, in which Neural ISPs correct residual aberrations rather than requiring increasingly complex optics.
\end{itemize}

\section{Related Work}
\label{sec:related}

\paragraph{Learned ISPs and end-to-end image restoration.}
A growing body of work has explored replacing portions of the classical ISP pipeline with learned models. Gharbi et al.~\cite{gharbi2016demosaic} demonstrated that joint deep demosaicking and denoising substantially outperforms sequential classical stages. Chen et al.~\cite{chen2018sid} showed that a single end-to-end network can map extreme low-light RAW captures to clean RGB outputs. DeepISP~\cite{schwartz2018deepisp} and PyNET~\cite{ignatov2020pynet} extended this paradigm to the full RAW-to-RGB pipeline, replacing the conventional ISP with a single learned model. Heide et al.~\cite{heide2014flexisp} introduced FlexISP, a unified inverse-problem formulation for image processing tasks that anticipated many of these directions. These models are typically trained on conventional Bayer data and target generic image quality rather than a specific lens. Our work builds on this line of research but focuses on the scaling behavior of learned ISPs as pixel pitch shrinks and optical aberrations come to dominate the degradation budget, with the network tailored to a specific camera module.

\paragraph{PSF-aware restoration and optical aberration correction.}
Several prior works address spatially varying optical aberrations directly. Classical non-blind deconvolution can invert a known PSF but amplifies noise and treats deblurring as a stage separate from demosaicing and denoising. Schuler et al.~\cite{schuler2011nonstationary} introduced non-stationary correction of optical aberrations using measured or estimated PSFs, and more recent learning-based approaches~\cite{eboli2022aberration} invert lens blur with deep networks, though typically blindly and generically across lenses rather than per camera. Our approach differs by jointly handling demosaicing, denoising, and aberration correction in a single learned pass tailored to a specific camera module, rather than treating aberration correction as a separate stage.

\paragraph{Computational optics and end-to-end camera design.}
Sitzmann et al.~\cite{sitzmann2018endtoend} and Tseng et al.~\cite{tseng2021differentiable} pioneered end-to-end optimization of optical elements jointly with downstream image processing networks. This co-design philosophy aligns closely with the implications of our results: when the reconstruction network can absorb significant aberrations, the optical design space is meaningfully expanded.

\paragraph{Multi-frame and burst imaging.}
Mobile imaging has long relied on multi-frame fusion to overcome single-frame SNR limits. Hasinoff et al.~\cite{hasinoff2016burst} demonstrated the value of burst photography for HDR and low-light capture on mobile devices, and Wronski et al.~\cite{wronski2019handheld} extended this to super-resolution from handheld bursts. A subsequent line of learning-based models integrates alignment and fusion within the network: DBSR~\cite{bhat2021dbsr} aligns frames via learned optical flow, BIPNet~\cite{dudhane2022bipnet} fuses ``pseudo-burst'' features, Burstormer~\cite{dudhane2023burstormer} performs transformer-based deformable alignment, and BurstM~\cite{kang2024burstm} couples optical-flow alignment with a Fourier-space implicit model for multi-scale super-resolution. These methods perform strongly on standard burst benchmarks, but are trained on conventional 2$\times$2 Bayer data and generic degradations and target denoising or super-resolution rather than correction of a specific lens's aberrations. We adopt Burstormer and BurstM as baselines (Section~\ref{sec:lowlight}); even after adapting their inputs to our Hex CFA and retraining a learnable front-end on our data, they trail our camera-specific model by a wide margin---indicating that joint, per-camera training on the known PSF, CFA, and noise, rather than multi-frame fusion per se, is the key differentiator. Our multi-frame extension applies this principle to the SNR challenges of small-pixel sensors.

\paragraph{Pixel scaling and sensor physics.}
The interplay between shrinking pixel size, diffraction, and per-pixel SNR has been studied extensively in the sensor physics literature~\cite{wong1996cmos,catrysse2005roadmap}. Our contribution is complementary: we characterize how reconstruction algorithms shift the practical resolution limits implied by physical constraints.

\section{Experiment Design: Scaling Pixel Size and Aperture Together}

\begin{figure}[t]
    \centering
    \includegraphics[width=\columnwidth]{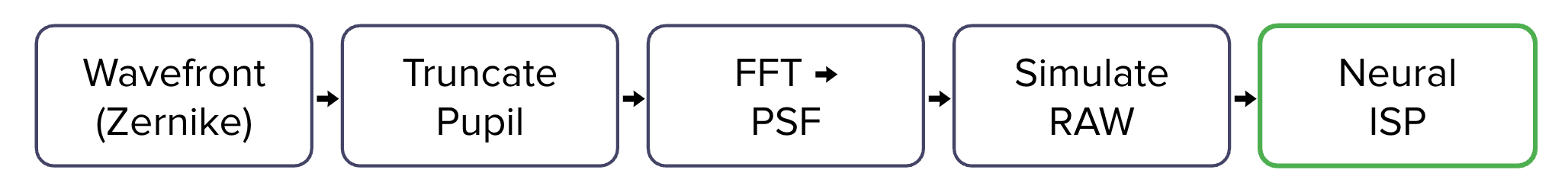}
    \caption{The simulation pipeline creates RAW images using the optics and sensor model, then restores them using either a traditional ISP or the Neural ISP.}
    \label{fig:fig3}
\end{figure}

\subsection{Simulating images from a set of cameras with realistic optical blur}

\begin{figure*}[t]
    \centering
    \includegraphics[width=0.75\textwidth]{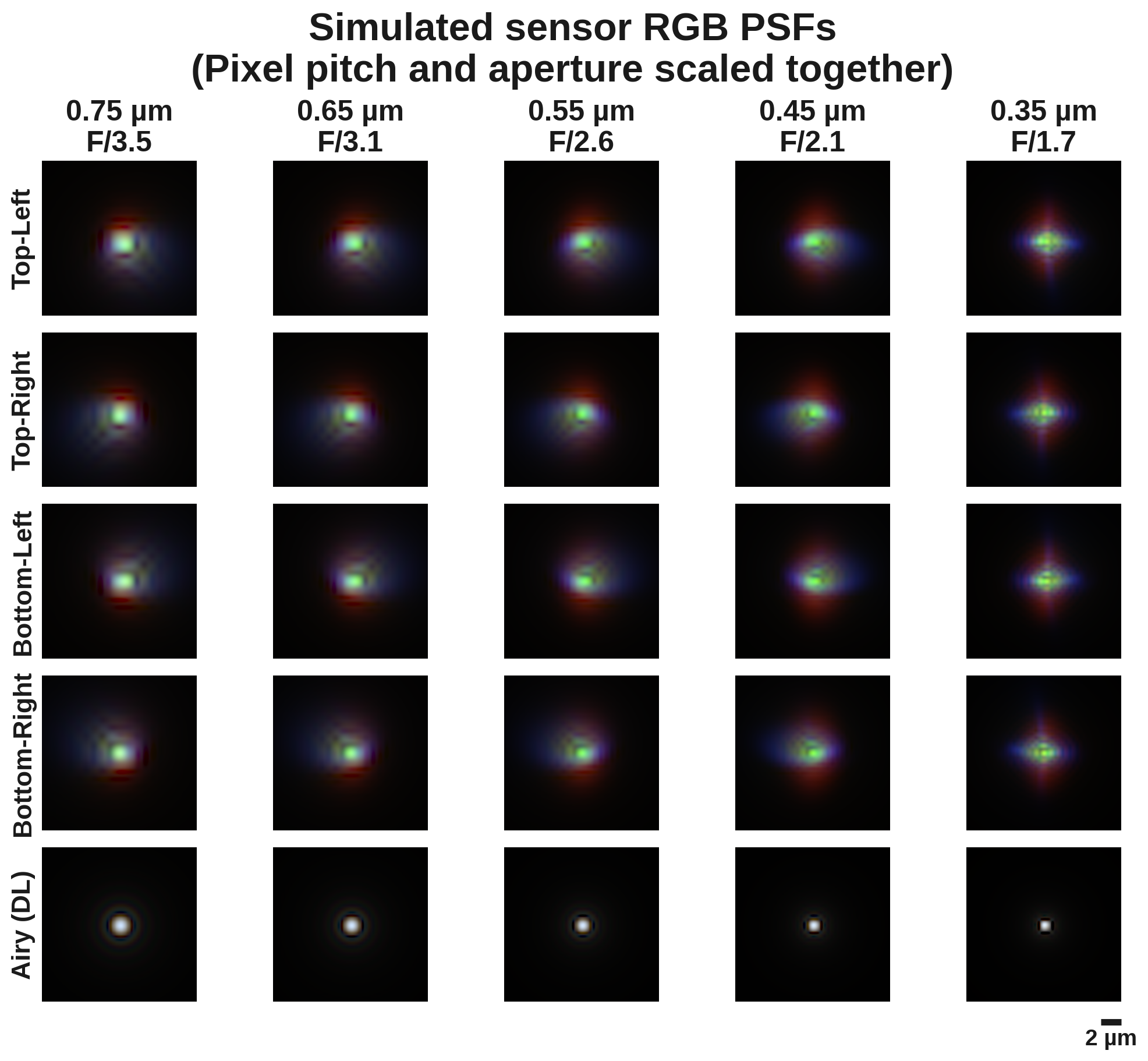}
    \caption{Simulated sensor RGB PSFs computed using the wavefronts at the exit pupil passing through the microlens (i.e., on-chip lens). The PSF maintains a similar overall spread despite the aperture increasing with smaller F/\#, due to more severe aberrations contributed by the marginal rays.}
    \label{fig:fig4}
\end{figure*}

The simulation pipeline (Figure~\ref{fig:fig3}) begins by modeling the optics and sensor behavior to produce realistic RAW images. We start with a representative Zernike wavefront for a typical smartphone telephoto lens module. The pupil (i.e., the aperture) is truncated based on the selected F/\# for each camera configuration. We then take a Fourier transform of the pupil function to propagate to the image sensor plane, and incorporate the sensor's on-chip lens (OCL) effects to obtain the final PSFs shown in Figure~\ref{fig:fig4}. The final RAW image is synthesized with Hex Bayer mosaicking (as found on many current 200MP sensors) and modeled sensor noise. We additionally account for SNR normalization across configurations. Further experimental details are provided in the appendix.

\subsection{Traditional ISP and Neural ISP image restoration methodology}

We feed the RAW image into a representative ``traditional ISP'' for comparison. It performs bilinear demosaicing, followed by a standard forward ISP color/tone chain (white balance, color correction, tone mapping, and gamma). Sensor noise is suppressed using a pyramid bilateral filter operating in YCbCr space, with conservative luma filtering to preserve detail and more aggressive chroma filtering to suppress the color noise introduced by demosaicing. A light unsharp mask is applied last to recover edge sharpness. This baseline is representative of a conventional hardware ISP without any learned or optics-aware processing.

We also feed the RAW image into our Neural ISP, which replaces the majority of the traditional imaging pipeline---including RAW denoising, CFA interpolation (demosaicing), multi-frame alignment and fusion, noise filtering, sharpening, and detail enhancement---with a single learned neural network tailored to each camera, operating end-to-end on the RAW sensor data. The network used for these experiments is a convolutional neural network (CNN) trained per camera module configuration, using paired RAW and ground truth data (typically obtained from real cameras through our automated optical and sensor characterization process, but in this case using our simulation data). Because the network is trained on data that captures the specific PSF, noise characteristics, and CFA layout of the target sensor and lens combination, it implicitly learns to invert these degradations jointly rather than treating them as independent processing stages. This is fundamentally different from a traditional ISP, where each stage operates without knowledge of the optical system that produced the input signal.

\paragraph{Network architecture.} The Neural ISP is a feed-forward, fully convolutional network that maps a single-channel mosaicked RAW frame (together with a per-pixel noise/exposure descriptor) to a full-resolution three-channel RGB image, performing demosaicing, denoising, and aberration correction jointly in a single pass without any upscaling. It comprises a lightweight demosaicing front-end followed by approximately six residual blocks with high-frequency attention gating, operating at roughly 128 feature channels (on the order of 45--50 convolutional layers and approximately 4.5~million parameters in the deployed form, requiring approximately 0.9~TMAC per megapixel of input). The residual blocks reparameterize into plain convolutions at inference, keeping the network efficient enough for on-device execution on mobile NPUs. Its effective receptive field (roughly 150--200~pixels) comfortably exceeds the point spread function support, allowing it to invert the optical blur. A multi-frame variant (Section~\ref{sec:lowlight}) shares the same backbone and additionally aligns the burst frames before fusing them. We train one network per camera configuration, holding the architecture and capacity fixed across all five pixel-pitch settings so that observed differences arise from the input data distribution rather than from architectural changes, for on the order of $10^5$ iterations using standard supervised image-restoration objectives.

The result is a full RAW-to-RGB Neural ISP that handles demosaicing, denoising, aberration correction, and detail recovery in a single forward pass. Standard post-processing steps such as color space conversion, tone mapping, and compression remain downstream and can be customized per application. The approach is agnostic to the specific CFA pattern and pixel size, and is designed for real-time on-device execution.

\subsection{Simulation configurations}

In practice, as pixel pitch decreases, camera module designers typically reduce the F-number proportionally. We formalize this as our experimental control variable: $\text{F/\#} \div \text{pitch} = \text{constant}$. This keeps the diffraction spot size fixed in pixel units across all configurations, so any changes in image quality are attributable to geometric aberrations and spatial sampling rather than diffraction.

We simulated five configurations spanning a 2$\times$ range of pixel pitch from 0.35 to 0.75~\um, as shown in Table~\ref{tab:simconfig}. For each configuration, the RAW images were synthesized using the forward model described above, and a neural network was trained on a set of more than 11{,}000 RAW/ground-truth pairs and evaluated on an unseen set of 159 test images to compute the metrics reported below.

\begin{table}[t]
\centering
\small
\caption{Simulation configurations for the proportional-scaling study. The F-number is scaled with pixel pitch so that $\text{F/\#} \div \text{pitch}$ is held constant ($\approx 4.7$). The aperture area then grows in inverse proportion to the pixel area, so the photon flux per pixel---and hence the shot-noise-limited SNR---remains constant across all five configurations. Aperture and pixel areas are relative to the 0.75~\um{} configuration.}
\label{tab:simconfig}
\begin{tabular}{ccccc}
\toprule
Pixel pitch & F/\# & Rel.\ aperture & Rel.\ pixel & Rel.\ flux \\
(\um) & & area & area & per pixel \\
\midrule
0.75 & 3.55 & 1.00$\times$ & 1.00$\times$ & 1.00$\times$ \\
0.65 & 3.07 & 1.34$\times$ & 0.75$\times$ & 1.00$\times$ \\
0.55 & 2.60 & 1.86$\times$ & 0.54$\times$ & 1.00$\times$ \\
0.45 & 2.13 & 2.78$\times$ & 0.36$\times$ & 1.00$\times$ \\
0.35 & 1.66 & 4.59$\times$ & 0.22$\times$ & 1.00$\times$ \\
\bottomrule
\end{tabular}
\end{table}

\section{Results: The Neural ISP Advantage Grows with Smaller Pixels}

Quantitative results across the five proportional-scaling configurations are summarized in Table~\ref{tab:mtf50} (MTF50) and Table~\ref{tab:perceptual} (perceptual metrics); the subsections below analyze each in turn.

\subsection{Visual comparison across pixel sizes shows Neural ISP recovers more detail}

Our initial results demonstrate the importance of Neural ISPs for image restoration as sensors move toward smaller pixels. We simulated several test images through our pipeline modeling the image formation and restoration process. For each test image, we show the input RAW (mosaicked), the traditional bilinear demosaic, the traditional ISP output, and the Neural ISP restoration across all five pixel pitch / F-number configurations.

\begin{figure*}[t]
    \centering
    \includegraphics[width=0.95\textwidth]{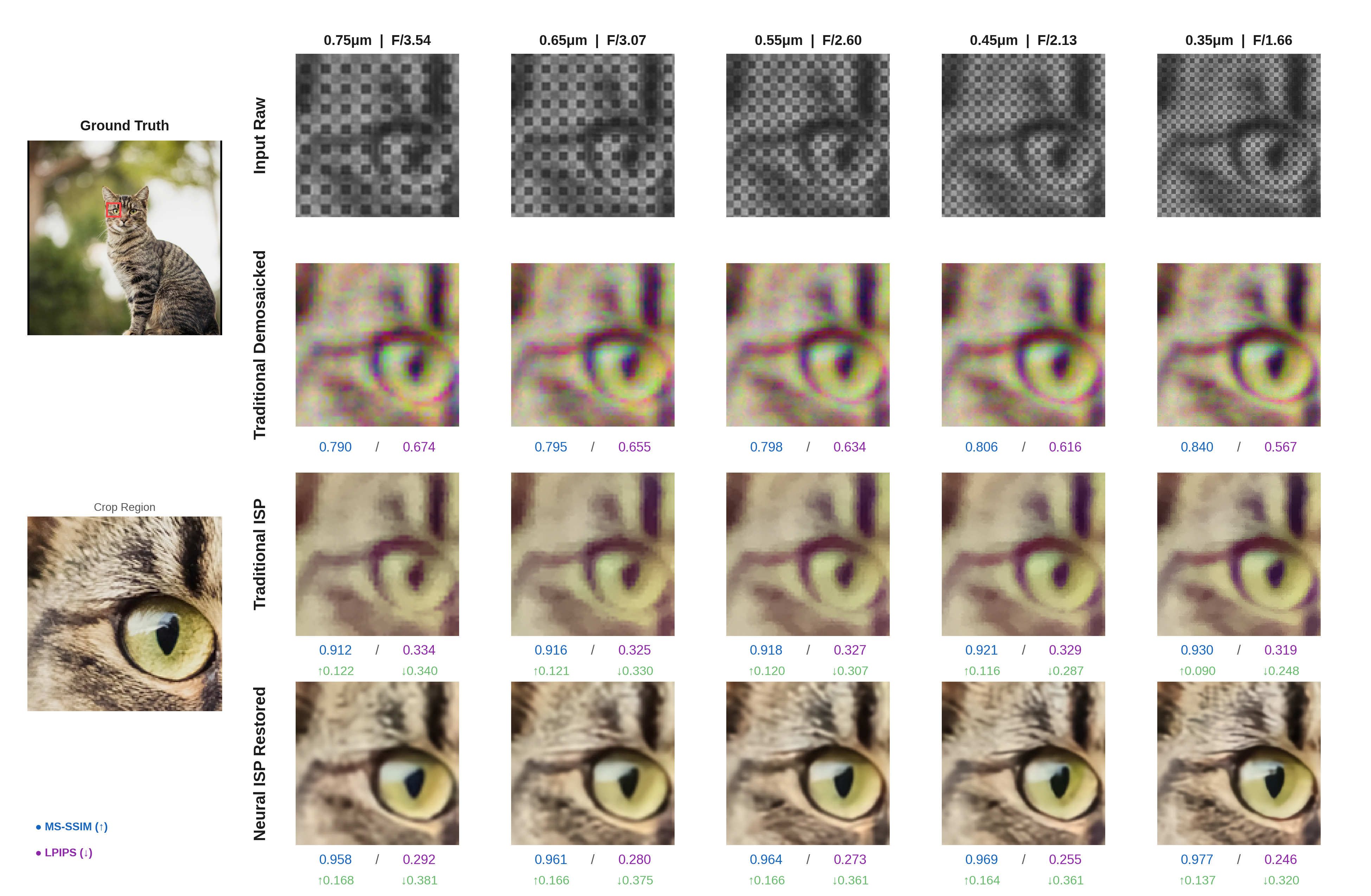}
    \caption{Sample natural images evaluated through the pipeline.}
    \label{fig:bright_natural}
\end{figure*}

\paragraph{Natural images.} On a detailed crop from natural images, the traditional ISP output remains blurry and over-smoothed even as pixel density increases (Figure~\ref{fig:bright_natural}). This is because the PSF also spans more pixels and consequently blurs the more densely sampled scene content. The Neural ISP output, by contrast, improves in sharpness and detail as pixel density increases because it is PSF-aware and can invert this effect. The 0.35~\um{} $|$ F/1.66 Neural ISP result resolves finer scene detail than the 0.75~\um{} $|$ F/3.55 result, despite the lens producing significantly worse aberrations at the smaller F/\#.

\begin{figure*}[t]
    \centering
    \includegraphics[width=0.95\textwidth]{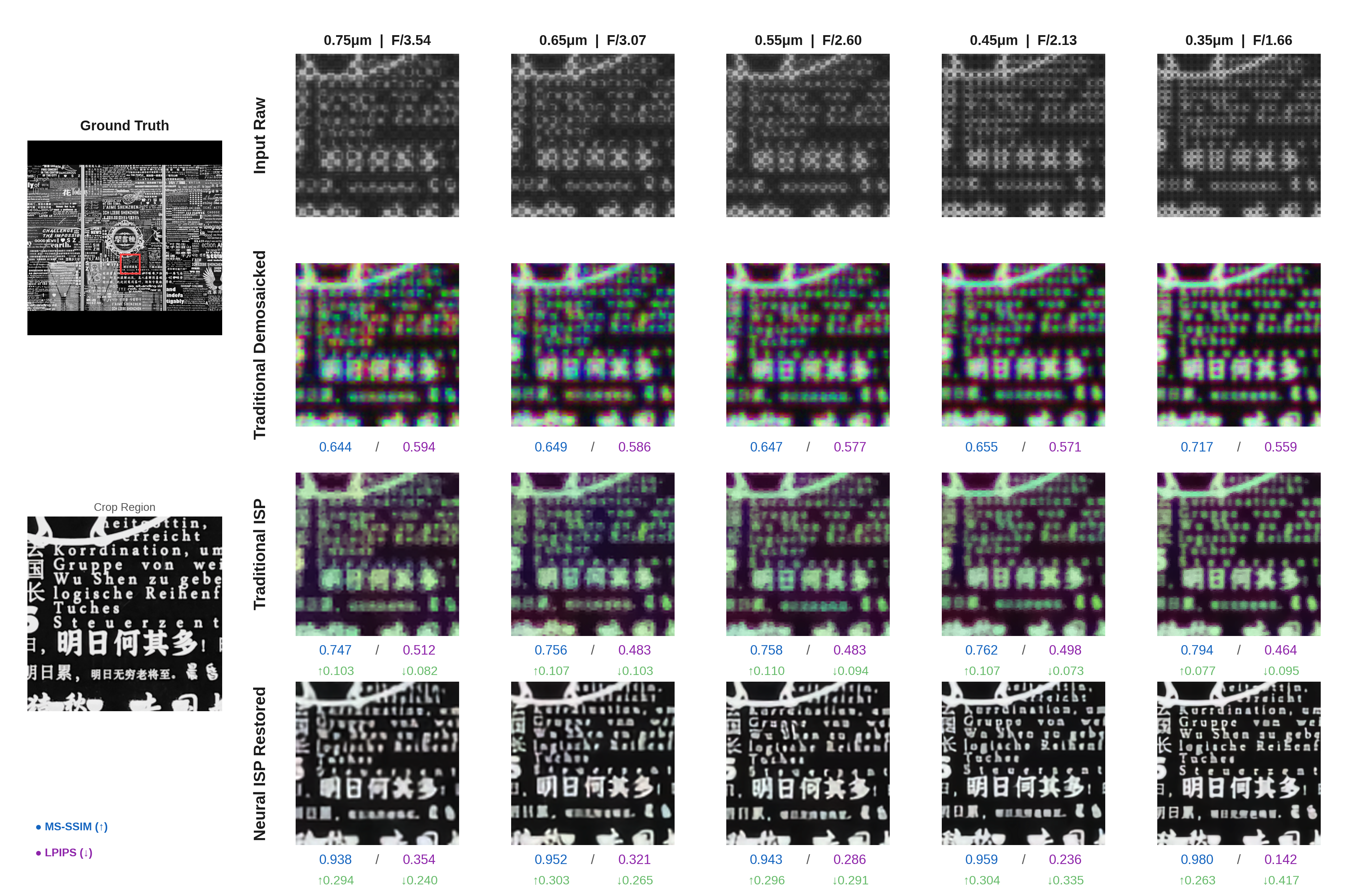}
    \caption{Sample text image evaluated through the pipeline.}
    \label{fig:bright_text}
\end{figure*}

\paragraph{Text and fine detail.} On a mixed-script text chart (Latin and CJK characters), the traditional ISP output is essentially unreadable across all configurations due to severe artifacts from Hex CFA demosaicing and uncorrected chromatic aberrations (Figure~\ref{fig:bright_text}). The Neural ISP recovers legible text across the full pitch range, with progressively sharper rendering at smaller pitches.

\begin{figure*}[t]
    \centering
    \includegraphics[width=0.95\textwidth]{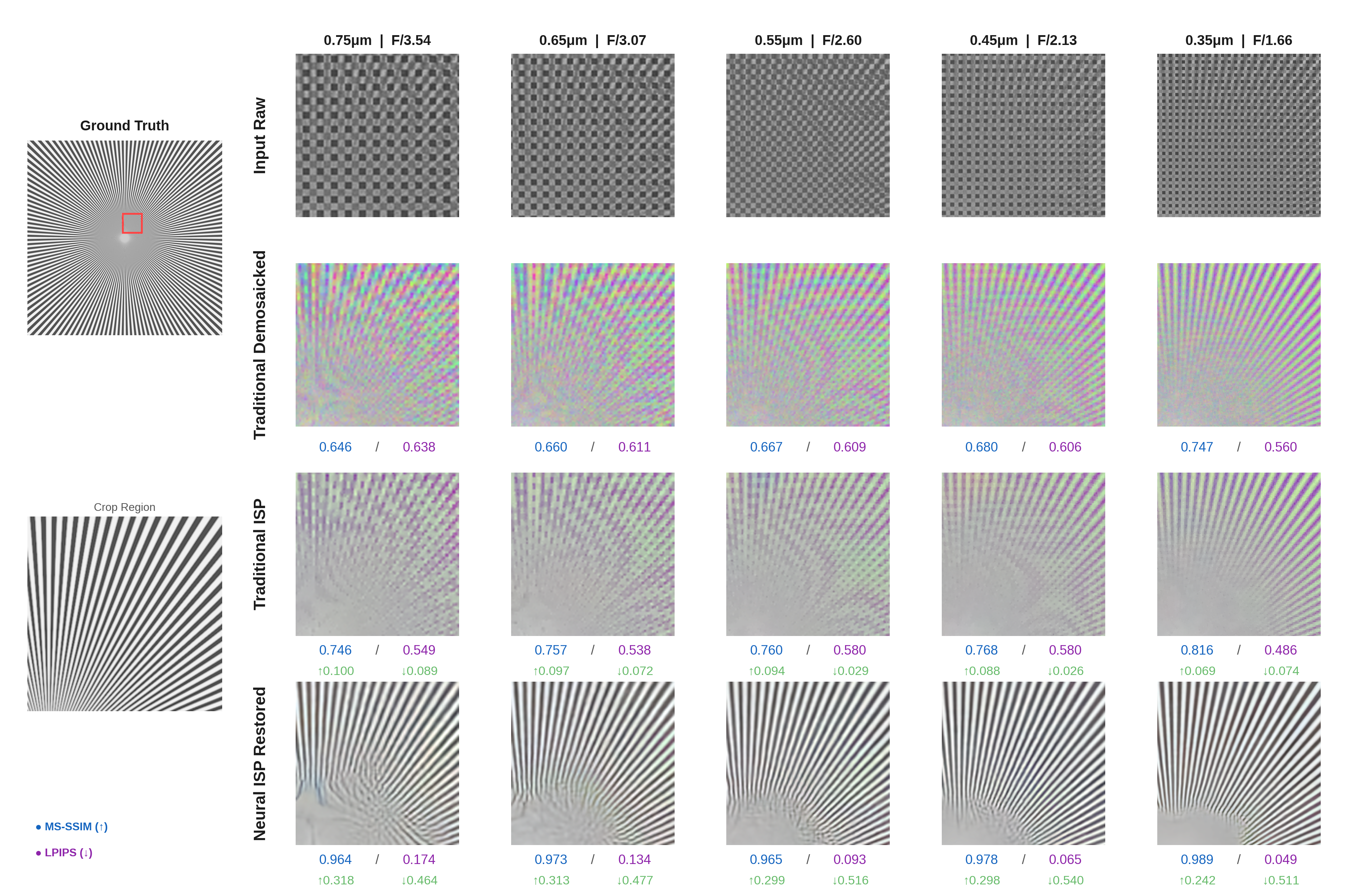}
    \caption{Sample resolution chart evaluated through the pipeline.}
    \label{fig:bright_resolution}
\end{figure*}

\paragraph{Resolution charts.} On converging line patterns, higher pixel density helps resolve finer lines in both cases (Figure~\ref{fig:bright_resolution}). However, the traditional ISP still exhibits moir\'e artifacts at smaller pixel sizes. The Neural ISP cleanly resolves the line patterns through PSF-aware deconvolution, with visibly finer detail preserved at 0.35~\um{} than at 0.75~\um.

\subsection{The MTF resolution analysis shows Neural ISP retains more spatial frequencies}

\begin{figure*}[t]
    \centering
    \includegraphics[width=0.95\textwidth]{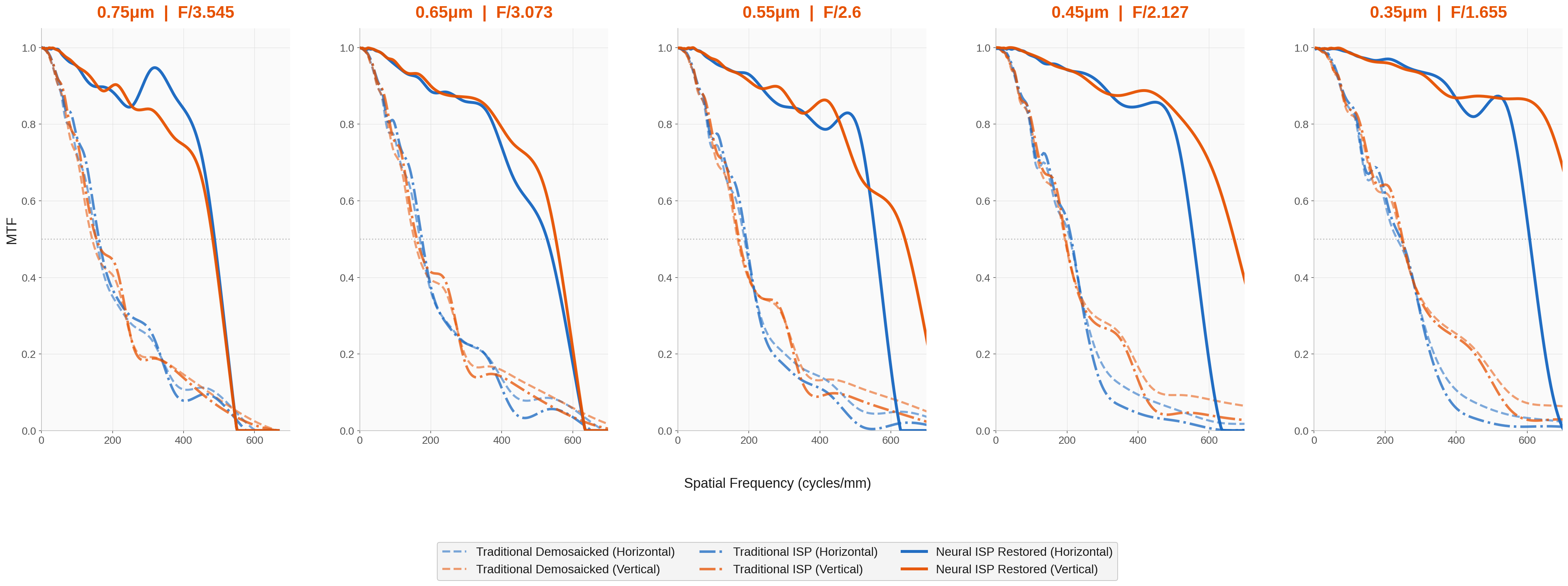}
    \caption{MTF curves for the different simulation cases show slower fall-off with Neural ISP.}
    \label{fig:mtf_bright}
\end{figure*}

The Modulation Transfer Function (MTF) curves (Figure~\ref{fig:mtf_bright}) quantify how the two imaging pipelines attenuate spatial frequencies and, consequently, detail. The ideal imaging system would have a flat MTF curve to maintain all spatial frequencies. For the traditional ISP, the MTF drops off rapidly at moderate spatial frequencies and degrades progressively as we move to smaller pixels and faster apertures. The traditional ISP behaves similarly to the demosaiced output because the simple operations in the pipeline do not expand the spatial frequency limit. For example, the unsharp mask (USM), a common step in traditional ISP, subtracts low-frequency content to enhance apparent sharpness but does not recover high-frequency content. For the Neural ISP, the MTF curves are elevated and maintain high contrast out to much higher frequencies, with the gap between the two methods widening at smaller pixel sizes.  We note that at the highest spatial frequencies, the Neural ISP output exhibits a sharper roll-off than the traditional demosaiced output. The network effectively trades some very high-frequency content (near the Nyquist limit) for robust mid-frequency recovery, a perceptually favorable exchange that is also visible in the strong anti-aliasing and moir\'e reduction observed in the resolution charts.

\begin{table*}[t]
\centering
\small
\caption{MTF50 (cycles/mm) for the proportional-scaling study (single-frame, 35~dB bright light); higher is better. Horizontal (H) and vertical (V) directions are reported separately---the H/V asymmetry is characteristic of the folded periscope optics, which produce different aberrations in each direction. The traditional ISP closely tracks the demosaiced output (its local operations do not extend the frequency limit), whereas the Neural ISP operates from a far higher baseline; the absolute Neural advantage increases as pixel pitch shrinks.}
\label{tab:mtf50}
\begin{tabular}{cc rrr rrr}
\toprule
& & \multicolumn{3}{c}{MTF50 horizontal} & \multicolumn{3}{c}{MTF50 vertical} \\
\cmidrule(lr){3-5}\cmidrule(lr){6-8}
Pitch (\um) & F/\# & Demosaiced & Traditional & Neural & Demosaiced & Traditional & Neural \\
\midrule
0.75 & 3.55 & 155.5 & 163.2 & 482.4 & 142.8 & 159.0 & 476.6 \\
0.65 & 3.07 & 170.3 & 176.7 & 527.3 & 155.6 & 162.0 & 546.6 \\
0.55 & 2.60 & 186.4 & 191.3 & 552.1 & 170.2 & 174.8 & 633.3 \\
0.45 & 2.13 & 206.4 & 209.8 & 553.3 & 193.4 & 195.9 & 668.5 \\
0.35 & 1.66 & 236.1 & 244.4 & 602.5 & 244.6 & 248.1 & 745.4 \\
\bottomrule
\end{tabular}
\end{table*}

The MTF50 metric measures the spatial frequency at which the attenuation falls to 50\% and serves as an indicator of system resolution.  MTF50 improves with both methods at smaller pixels (since there are more pixels per mm), but the Neural ISP improves from a far higher baseline---roughly 3$\times$ that of the traditional ISP across the entire sweep (Table~\ref{tab:mtf50}). At 0.35~\um{} the Neural ISP reaches 603~cycles/mm (horizontal) and 745~cycles/mm (vertical), versus 244 and 248 for the traditional ISP, a 2.5--3$\times$ advantage.

\subsection{Perceptual metrics for image quality show improved performance with Neural ISP at smaller pixels}

\begin{table*}[t]
\centering
\small
\caption{Perceptual metrics for the proportional-scaling study (single-frame, 35~dB bright light), averaged over the 159-image test set. MS-SSIM: higher is better; LPIPS: lower is better. The Neural ISP improves on both metrics as pixels shrink, whereas the traditional ISP improves only marginally on MS-SSIM and stays roughly flat on LPIPS.}
\label{tab:perceptual}
\begin{tabular}{cc rrr rrr}
\toprule
& & \multicolumn{3}{c}{MS-SSIM $\uparrow$} & \multicolumn{3}{c}{LPIPS $\downarrow$} \\
\cmidrule(lr){3-5}\cmidrule(lr){6-8}
Pitch (\um) & F/\# & Demosaiced & Traditional & Neural & Demosaiced & Traditional & Neural \\
\midrule
0.75 & 3.55 & 0.775 & 0.871 & 0.957 & 0.552 & 0.356 & 0.244 \\
0.65 & 3.07 & 0.780 & 0.877 & 0.963 & 0.543 & 0.345 & 0.223 \\
0.55 & 2.60 & 0.783 & 0.880 & 0.961 & 0.532 & 0.347 & 0.204 \\
0.45 & 2.13 & 0.792 & 0.886 & 0.968 & 0.520 & 0.349 & 0.183 \\
0.35 & 1.66 & 0.826 & 0.901 & 0.979 & 0.485 & 0.321 & 0.151 \\
\bottomrule
\end{tabular}
\end{table*}

\begin{itemize}[leftmargin=*]
    \item \textbf{MS-SSIM:} For the Multi-Scale Structural Similarity Metric, the Neural ISP scores are very high across the board (0.957 to 0.979), and the absolute delta decreases slightly at the smallest pitches. This is a known ceiling effect: MS-SSIM compresses gains as scores approach 1.0, making each incremental improvement harder to register. The traditional ISP improves from 0.871 to 0.901, reflecting the higher sampling density.

    \item \textbf{LPIPS:} Often regarded as more perceptually accurate than MS-SSIM, the Learned Perceptual Image Patch Similarity score for the Neural ISP improves from 0.244 at 0.75~\um{} to 0.151 at 0.35~\um, while the traditional ISP shows relatively flat perceptual quality (0.356 to 0.321). The traditional ISP results reflect the observations from the natural-image analysis: increased sampling density is offset by a larger effective PSF, leading to persistent blurriness even at reduced sensor pixel sizes.
\end{itemize}

\subsection{Robustness across aperture conditions}

We also performed an ablation study by shrinking the pixel pitch while keeping the aperture fixed at F/2.6. In this scenario, the diffraction spot covers progressively more pixels as pitch decreases, and per-pixel SNR drops because the same total light is spread across more, smaller pixels. This represents the case of keeping the optics the same while moving to smaller pixel sensors.  As expected, the traditional ISP degrades more severely than in the proportional-scaling case, confirming the well-known ``diffraction wall'' challenge for small pixels. Critically, the Neural ISP remained robust, showing relative insensitivity to whether image degradation originated from geometric aberrations (fast aperture) or diffraction (slow aperture). See the appendix for additional figures.

While the Neural ISP handles both effectively, proportional scaling of F/\# to pitch remains clearly preferable, as it maintains per-pixel SNR, avoids the diffraction wall, and provides the network with the cleanest spatial information from which to maximize resolution.

\subsection{Low-light performance and multi-frame Neural ISP}
\label{sec:lowlight}

A common concern with small pixels is low-light performance, since reduced well capacity decreases the obtainable shot-noise-limited SNR.  

We evaluated this by comparing the traditional ISP and Neural ISP at approximately 15~dB SNR---a challenging scenario equivalent to dark indoor or twilight conditions---and at 35~dB SNR representative of bright-light capture (Figure~\ref{fig:mf_visual}). In the single-frame case, the Neural ISP denoises more effectively than the traditional method. The trade-off at low light is some loss of detail and additional smoothing compared to the 35~dB case (while still preserving more detail than the traditional ISP). This behavior is expected, as the network is trained to avoid hallucinating features beyond what detectable signal is present in the single-frame input RAW.

However, the single-frame 15~dB case represents a worst-case scenario. In practice, both traditional and Neural ISP pipelines employ multi-frame fusion to improve effective SNR. We therefore extend our analysis to a multi-frame Neural ISP that ingests a burst of RAW frames and produces a single restored RGB output. We focus on the most challenging configuration in our study, 0.35~\um{} pixel pitch at F/1.66, where the combination of small pixels, severe aberrations, and low SNR places the greatest burden on the restoration pipeline. Bursts of 10 frames at 15~dB per-frame SNR are processed by both the traditional ISP (with multi-frame averaging-based fusion) and the multi-frame Neural ISP.

\definecolor{metblue}{HTML}{1F77B4}
\definecolor{metpurple}{HTML}{9C27B0}
\newcommand{\cellw}{0.108\textwidth}
\newcommand{\mfcol}[1]{\parbox[b]{\cellw}{\centering\scriptsize #1\strut}}
\newcommand{\mfe}{\makebox[\cellw]{}}
\begin{figure*}[tp]
    \centering
    \footnotesize
    \setlength{\tabcolsep}{1pt}
    \renewcommand{\arraystretch}{0}
    \begin{tabular}{cccccc}
        \mfcol{SF \\ Traditional ISP} & \mfcol{SF \\ Neural ISP} & \mfcol{MF \\ Traditional ISP} & \mfcol{MF \\ Neural ISP} & \mfcol{MF Burstormer \\ (off-the-shelf)} & \mfcol{MF BurstM \\ (off-the-shelf)} \\[2pt]
        \shortstack{\includegraphics[width=\cellw]{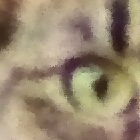} \\ \scriptsize {\color{metblue}0.881}\,/\,{\color{metpurple}0.546}} & \shortstack{\includegraphics[width=\cellw]{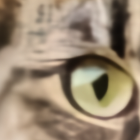} \\ \scriptsize {\color{metblue}0.952}\,/\,{\color{metpurple}0.304}} & \shortstack{\includegraphics[width=\cellw]{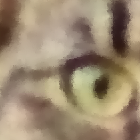} \\ \scriptsize {\color{metblue}0.898}\,/\,{\color{metpurple}0.496}} & \shortstack{\includegraphics[width=\cellw]{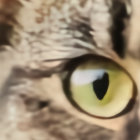} \\ \scriptsize {\color{metblue}0.969}\,/\,{\color{metpurple}0.270}} & \shortstack{\includegraphics[width=\cellw]{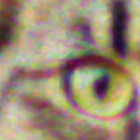} \\ \scriptsize {\color{metblue}0.741}\,/\,{\color{metpurple}0.756}} & \shortstack{\includegraphics[width=\cellw]{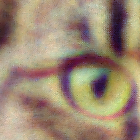} \\ \scriptsize {\color{metblue}0.891}\,/\,{\color{metpurple}0.458}} \\[2pt]
        \shortstack{\includegraphics[width=\cellw]{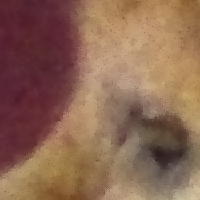} \\ \scriptsize {\color{metblue}0.943}\,/\,{\color{metpurple}0.352}} & \shortstack{\includegraphics[width=\cellw]{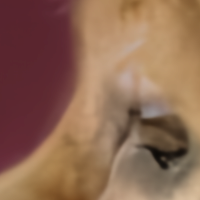} \\ \scriptsize {\color{metblue}0.969}\,/\,{\color{metpurple}0.233}} & \shortstack{\includegraphics[width=\cellw]{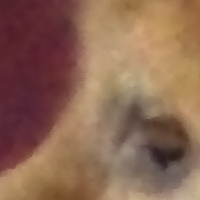} \\ \scriptsize {\color{metblue}0.948}\,/\,{\color{metpurple}0.354}} & \shortstack{\includegraphics[width=\cellw]{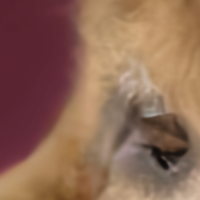} \\ \scriptsize {\color{metblue}0.971}\,/\,{\color{metpurple}0.219}} & \shortstack{\includegraphics[width=\cellw]{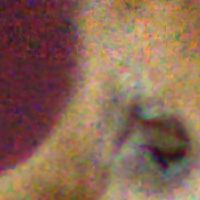} \\ \scriptsize {\color{metblue}0.802}\,/\,{\color{metpurple}0.654}} & \shortstack{\includegraphics[width=\cellw]{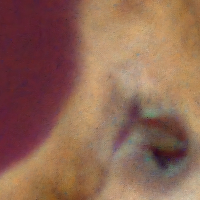} \\ \scriptsize {\color{metblue}0.942}\,/\,{\color{metpurple}0.310}} \\[2pt]
        \shortstack{\includegraphics[width=\cellw]{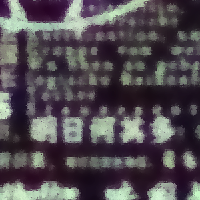} \\ \scriptsize {\color{metblue}0.807}\,/\,{\color{metpurple}0.462}} & \shortstack{\includegraphics[width=\cellw]{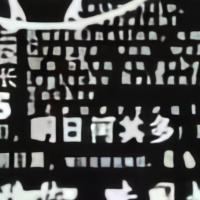} \\ \scriptsize {\color{metblue}0.949}\,/\,{\color{metpurple}0.261}} & \shortstack{\includegraphics[width=\cellw]{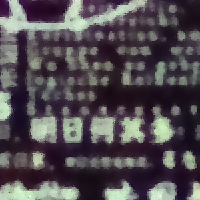} \\ \scriptsize {\color{metblue}0.830}\,/\,{\color{metpurple}0.407}} & \shortstack{\includegraphics[width=\cellw]{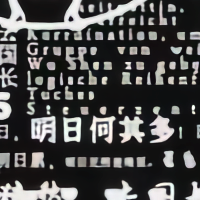} \\ \scriptsize {\color{metblue}0.976}\,/\,{\color{metpurple}0.160}} & \shortstack{\includegraphics[width=\cellw]{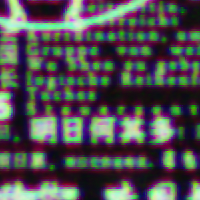} \\ \scriptsize {\color{metblue}0.711}\,/\,{\color{metpurple}0.690}} & \shortstack{\includegraphics[width=\cellw]{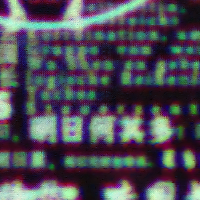} \\ \scriptsize {\color{metblue}0.744}\,/\,{\color{metpurple}0.538}} \\[2pt]
        \shortstack{\includegraphics[width=\cellw]{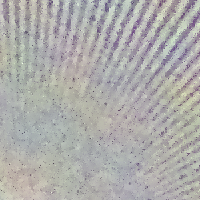} \\ \scriptsize {\color{metblue}0.839}\,/\,{\color{metpurple}0.596}} & \shortstack{\includegraphics[width=\cellw]{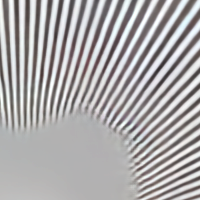} \\ \scriptsize {\color{metblue}0.983}\,/\,{\color{metpurple}0.060}} & \shortstack{\includegraphics[width=\cellw]{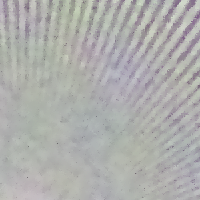} \\ \scriptsize {\color{metblue}0.845}\,/\,{\color{metpurple}0.488}} & \shortstack{\includegraphics[width=\cellw]{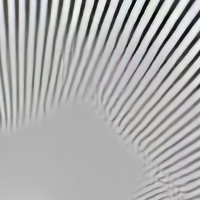} \\ \scriptsize {\color{metblue}0.961}\,/\,{\color{metpurple}0.151}} & \shortstack{\includegraphics[width=\cellw]{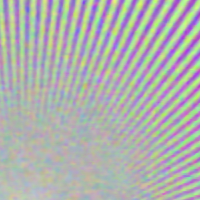} \\ \scriptsize {\color{metblue}0.745}\,/\,{\color{metpurple}0.569}} & \shortstack{\includegraphics[width=\cellw]{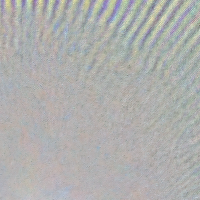} \\ \scriptsize {\color{metblue}0.737}\,/\,{\color{metpurple}0.571}} \\[2pt]
    \end{tabular}
    \\[6pt]
    \begin{tabular}{cccccc}
        \mfcol{MF Burstormer \\ (retrained)} & \mfcol{MF BurstM \\ (retrained)} & \mfcol{SF Trad ISP \\ (35\,dB ref)} & \mfcol{SF Neural ISP \\ (35\,dB ref)} & \mfcol{Ground \\ Truth} & \mfe \\[2pt]
        \shortstack{\includegraphics[width=\cellw]{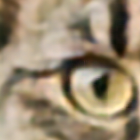} \\ \scriptsize {\color{metblue}0.479}\,/\,{\color{metpurple}0.730}} & \shortstack{\includegraphics[width=\cellw]{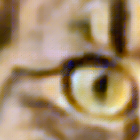} \\ \scriptsize {\color{metblue}0.421}\,/\,{\color{metpurple}0.671}} & \shortstack{\includegraphics[width=\cellw]{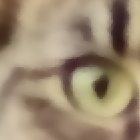} \\ \scriptsize {\color{metblue}0.919}\,/\,{\color{metpurple}0.376}} & \shortstack{\includegraphics[width=\cellw]{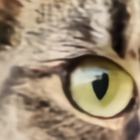} \\ \scriptsize {\color{metblue}0.975}\,/\,{\color{metpurple}0.252}} & \shortstack{\includegraphics[width=\cellw]{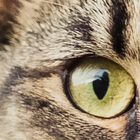}\\[1pt] {\scriptsize\phantom{0.000\,/\,0.000}}} & \mfe \\[2pt]
        \shortstack{\includegraphics[width=\cellw]{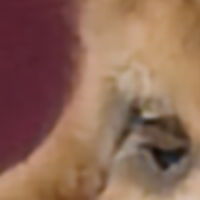} \\ \scriptsize {\color{metblue}0.909}\,/\,{\color{metpurple}0.488}} & \shortstack{\includegraphics[width=\cellw]{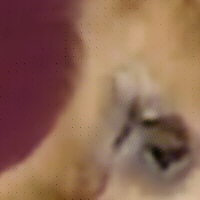} \\ \scriptsize {\color{metblue}0.894}\,/\,{\color{metpurple}0.462}} & \shortstack{\includegraphics[width=\cellw]{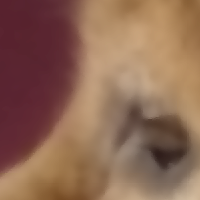} \\ \scriptsize {\color{metblue}0.966}\,/\,{\color{metpurple}0.281}} & \shortstack{\includegraphics[width=\cellw]{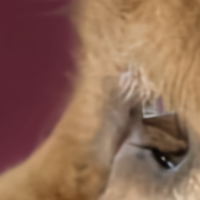} \\ \scriptsize {\color{metblue}0.985}\,/\,{\color{metpurple}0.162}} & \shortstack{\includegraphics[width=\cellw]{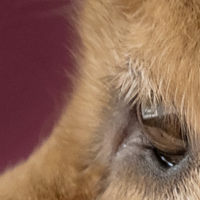}\\[1pt] {\scriptsize\phantom{0.000\,/\,0.000}}} & \mfe \\[2pt]
        \shortstack{\includegraphics[width=\cellw]{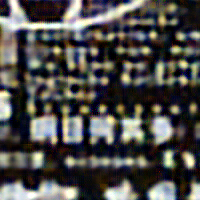} \\ \scriptsize {\color{metblue}0.796}\,/\,{\color{metpurple}0.743}} & \shortstack{\includegraphics[width=\cellw]{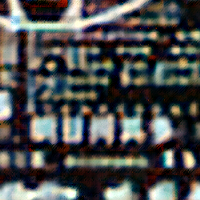} \\ \scriptsize {\color{metblue}0.766}\,/\,{\color{metpurple}0.719}} & \shortstack{\includegraphics[width=\cellw]{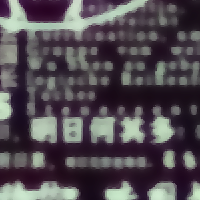} \\ \scriptsize {\color{metblue}0.825}\,/\,{\color{metpurple}0.406}} & \shortstack{\includegraphics[width=\cellw]{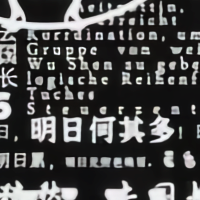} \\ \scriptsize {\color{metblue}0.979}\,/\,{\color{metpurple}0.147}} & \shortstack{\includegraphics[width=\cellw]{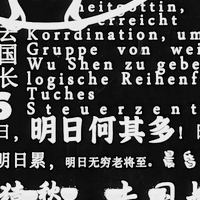}\\[1pt] {\scriptsize\phantom{0.000\,/\,0.000}}} & \mfe \\[2pt]
        \shortstack{\includegraphics[width=\cellw]{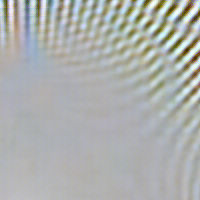} \\ \scriptsize {\color{metblue}0.433}\,/\,{\color{metpurple}0.786}} & \shortstack{\includegraphics[width=\cellw]{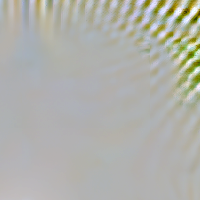} \\ \scriptsize {\color{metblue}0.322}\,/\,{\color{metpurple}0.834}} & \shortstack{\includegraphics[width=\cellw]{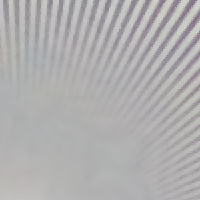} \\ \scriptsize {\color{metblue}0.882}\,/\,{\color{metpurple}0.342}} & \shortstack{\includegraphics[width=\cellw]{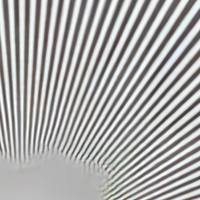} \\ \scriptsize {\color{metblue}0.989}\,/\,{\color{metpurple}0.048}} & \shortstack{\includegraphics[width=\cellw]{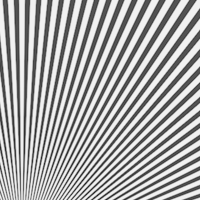}\\[1pt] {\scriptsize\phantom{0.000\,/\,0.000}}} & \mfe \\[2pt]
    \end{tabular}
    \vspace{-0.4em}
    \caption{Visual comparison at 0.35~\um, F/1.66 under low-SNR conditions (15~dB per-frame, 10-frame burst), evaluated on four representative crops (rows: cat eye, dog eye, mixed-script text chart, converging lines). \textbf{Top panel}: the four ``established'' pipelines plus the two off-the-shelf burst networks. The multi-frame Neural ISP recovers detail that the traditional pipelines and off-the-shelf burst networks miss. \textbf{Bottom panel}: the two retrained burst-network adapters (frozen backbone + small Hex CFA front-end), the bright-light (35~dB) single-frame references, and ground truth. Retraining the burst networks on our paired data closes part of the gap to ground truth but stays well below the camera-specific Neural ISP and the bright-light reference. Numbers under each cell are MS-SSIM~$\uparrow$ ({\color{metblue}blue}) / LPIPS~$\downarrow$ ({\color{metpurple}purple}); per-crop values vary substantially with image content, so quantitative conclusions are drawn from the 159-image averages in Table~\ref{tab:multiframe} rather than from individual crops. All crops are brightness-boosted ($\times 1.5$) for visibility; the brightness factor is identical across all cells so they are directly comparable.}
    \label{fig:mf_visual}
\end{figure*}

\begin{figure}[t]
    \centering
    \includegraphics[width=\columnwidth]{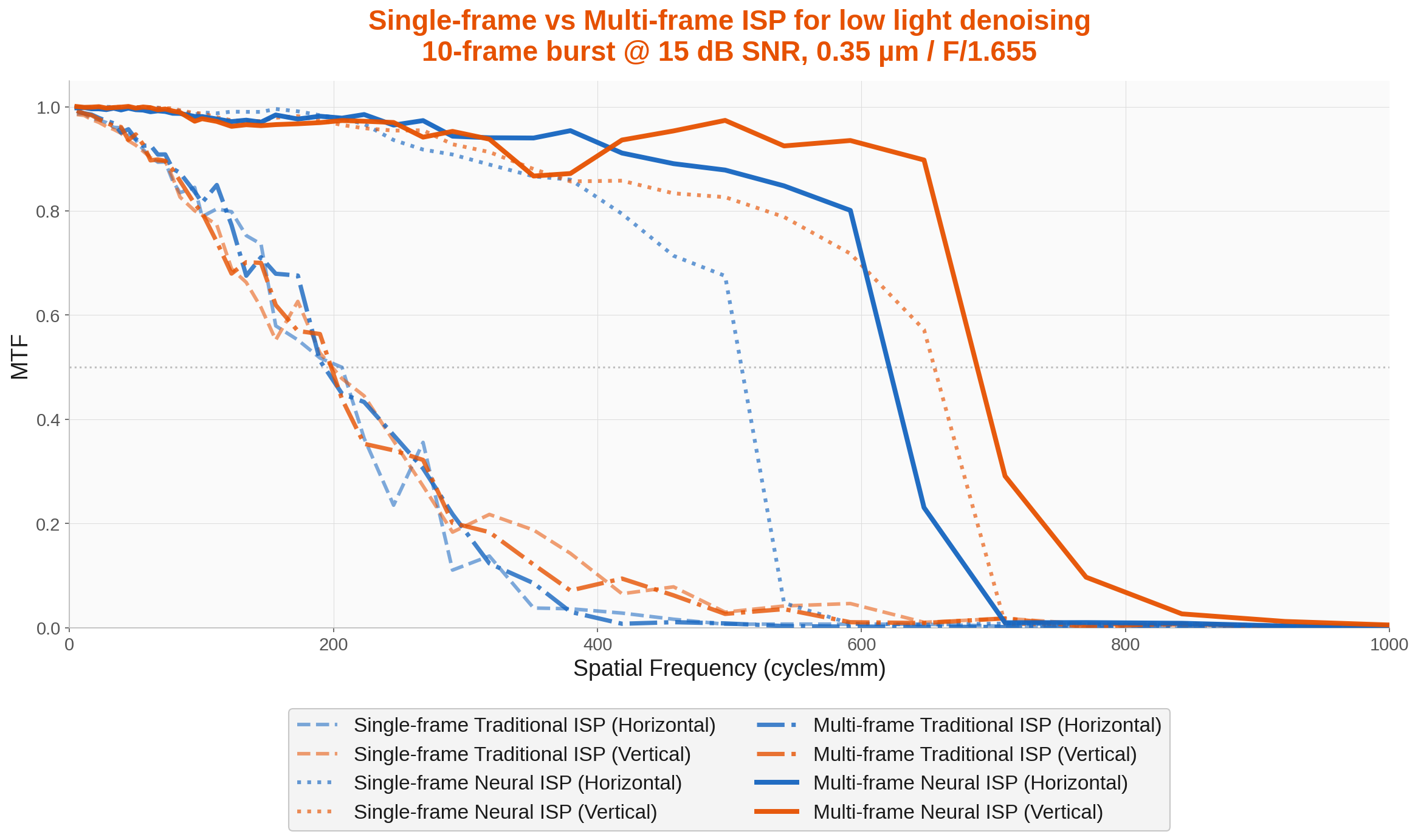}
    \caption{MTF curves at 0.35~\um, F/1.66 comparing single-frame and multi-frame configurations under 15~dB per-frame SNR. The multi-frame Neural ISP preserves high contrast at substantially higher spatial frequencies than the single-frame Neural ISP, while the traditional ISP curves are nearly indistinguishable between single- and multi-frame variants.}
    \label{fig:mf_mtf_curves}
\end{figure}

\begin{table*}[t]
\centering
\footnotesize
\caption{Multi-frame low-light results at 0.35~\um, F/1.66 under 15~dB per-frame SNR (10-frame burst), averaged over 159 test images. SF/MF denote single-/multi-frame; MTF50 is in cycles/mm; compute is MACs per output megapixel. The Neural ISP is trained on the target sensor and optics; Burstormer~\cite{dudhane2023burstormer} and BurstM~\cite{kang2024burstm} are recent third-party burst networks. Both backbones were originally trained on standard 2$\times$2 Bayer CFA inputs (RGB bursts for Burstormer, packed RGGB for BurstM) and so cannot consume our 8$\times$8 Hex CFA mosaic directly. The ``off-the-shelf'' rows feed each backbone its expected input via a fixed front-end---a Hex-aware bilinear demosaic to RGB for Burstormer, and a Hex-to-packed-Bayer reordering for BurstM---with the published weights used unchanged. The ``retrained'' rows replace this fixed front-end with a small learnable Hex CFA adapter (PixelUnshuffle by the 8-pixel period, three $3{\times}3$ convolutions at 64 channels, then a PixelShuffle to the backbone's expected layout) trained on our paired data while the backbone stays frozen at the published weights; this adds $\sim$0.22~M trainable parameters for Burstormer and $\sim$0.15~M for BurstM (rounded into the Params column), with negligible compute overhead. $^{*}$BurstM's compute is its native-output cost and includes ${\sim}16\times$ overhead from an internal 4$\times$ super-resolution grid (SR-normalized ${\approx}2.9$~TMAC/MP). Off-the-shelf, both burst networks track the traditional ISP in quality; retraining the Hex CFA adapter yields a consistent uplift in the averaged metrics but closes only a small part of the gap to the camera-specific Neural ISP, despite costing far more compute.}
\label{tab:multiframe}
\begin{tabular}{l c c ccc cc}
\toprule
Method & Params (M) & TMAC/MP & PSNR (dB) $\uparrow$ & MS-SSIM $\uparrow$ & LPIPS $\downarrow$ & MTF50-H $\uparrow$ & MTF50-V $\uparrow$ \\
\midrule
SF Traditional ISP & --- & --- & 25.54 & 0.864 & 0.482 & 206.2 & 199.0 \\
SF Neural ISP      & 4.5 & 0.9 & 29.91 & 0.954 & 0.225 & 509.1 & 655.1 \\
MF Traditional ISP & --- & --- & 25.99 & 0.878 & 0.414 & 193.1 & 198.3 \\
MF Neural ISP      & 4.7 & 1.1 & 30.89 & 0.972 & 0.171 & 621.0 & 687.6 \\
\midrule
MF Burstormer (off-the-shelf) & 3.1 & 15.5 & 24.49 & 0.800 & 0.645 & 222.0 & 237.9 \\
MF BurstM (off-the-shelf)     & 14.0 & 47$^{*}$ & 25.25 & 0.859 & 0.437 & 217.2 & 245.3 \\
MF Burstormer (retrained)     & 3.4 & 15.5 & 26.98 & 0.923 & 0.368 & 275.3 & 285.4 \\
MF BurstM (retrained)         & 14.2 & 47$^{*}$ & 26.29 & 0.910 & 0.374 & 241.4 & 254.4 \\
\bottomrule
\end{tabular}
\end{table*}

The results show a striking asymmetry between the traditional and Neural ISP pipelines (Table~\ref{tab:multiframe}). Multi-frame fusion improves the Neural ISP MTF50 from 509 to 621~cycles/mm horizontally ($+22\%$) and from 655 to 688~cycles/mm vertically ($+5\%$), and likewise improves its perceptual quality (MS-SSIM 0.954 to 0.972, LPIPS 0.225 to 0.171). This brings low-SNR performance close to the bright-light (35~dB) single-frame Neural ISP baseline of 603/745~cycles/mm reported in Table~\ref{tab:mtf50}---and in the horizontal direction it matches and slightly exceeds that baseline. In contrast, the traditional ISP barely changes: its MTF50 moves from 206 to 193~cycles/mm horizontally and from 199 to 198 vertically (no meaningful gain), with only a small perceptual improvement. This is consistent with the interpretation that the traditional ISP's resolution at small pixel sizes is limited by uncorrected PSF blur rather than by sensor noise: reducing noise via temporal averaging does not address the dominant degradation. The Neural ISP, by virtue of its joint model of PSF, demosaicing, and noise, converts the additional SNR margin from multi-frame fusion directly into recovered spatial frequencies.

For reference, Table~\ref{tab:multiframe} also includes two recent third-party multi-frame networks---Burstormer~\cite{dudhane2023burstormer}, a transformer-based burst restoration model, and BurstM~\cite{kang2024burstm}, a burst super-resolution model. Both networks were originally trained on 2$\times$2 Bayer CFA data, not the 8$\times$8 Hex CFA used here, so they must be supplied an input they can interpret. We evaluate each in two configurations. In the \emph{off-the-shelf} configuration the published weights are used unchanged: Hex CFA bursts are converted into the format each backbone expects via a fixed, hand-designed front-end---a Hex-aware bilinear demosaic to RGB for Burstormer, and a Hex-to-packed-Bayer reordering for BurstM---and the resulting RGB or packed-Bayer bursts are fed to the frozen network. In the \emph{retrained} configuration the backbone is again frozen at its published weights, but the fixed front-end is replaced by a small learnable \emph{Hex CFA adapter} (a PixelUnshuffle by the 8-pixel Hex period, three $3{\times}3$ convolutions at 64 channels, and a PixelShuffle back to whatever spatial layout the backbone expects: RGB at full resolution for Burstormer, packed RGGB at half resolution for BurstM). Only the adapter is optimized---on the same paired Hex CFA bursts we use to train the Neural ISP---so the comparison isolates the effect of giving each architecture a tailored input rather than expanding its capacity (the adapter adds $\sim$0.22~M parameters for Burstormer and $\sim$0.15~M for BurstM, less than 5\% of either backbone). Applied off-the-shelf neither network transfers to this regime: both remain near the traditional-ISP MTF50 ($\sim$220--245~cycles/mm) and below it perceptually, far short of the camera-specific Neural ISP, even though Burstormer is smaller (3.1~M parameters) and BurstM substantially larger (14.0~M) than our network (4.5~M). Retraining the adapter on our paired Hex data yields a consistent uplift across both architectures---PSNR +2.5~dB and +1.0~dB, MS-SSIM +0.12 and +0.05, LPIPS $-0.28$ and $-0.06$ for Burstormer and BurstM respectively, with MTF50 climbing into the 240--285~cycles/mm range. These retrained rows close only a small part of the gap to the camera-specific Neural ISP and still trail it by a wide margin in every metric, despite the BurstM backbone having three times the parameter count of our network. This indicates that the gains stem from camera-specific joint training on the known degradations rather than from model capacity. Fitting only a Hex CFA front-end leaves substantial joint denoising/deblurring performance on the table. 

These findings reinforce a broader theme of this work: traditional and Neural ISPs are bottlenecked by different effects at small pixel sizes, and improvements to inputs (e.g., more frames, better optics) yield very different returns for each. We leave a more detailed analysis of burst length, motion robustness, device-to-device variation, scaling across all five pixel pitch configurations, and adaptive frame weighting to future work.

\section{Conclusion: What This Means for the Next Generation of Image Sensors}

\subsection{Neural ISPs enable smaller pixels}

Traditionally, pixel scaling has faced several challenges: smaller pixels imply higher noise and demand better (and more expensive) optics, because traditional ISPs cannot incorporate PSF effects or correct geometric aberrations. Our results suggest that with Neural ISP--based image restoration, these challenges can be overcome.

The key result is that the denser spatial sampling provided by smaller pixels is a net positive, even when accompanied by worse geometric aberrations, provided that:

\begin{itemize}[leftmargin=*]
    \item Ideally, the aperture is scaled proportionally to maintain per-pixel SNR.
    \item A Neural ISP is used to jointly handle demosaicing, denoising, and optical aberration correction.
\end{itemize}

\subsection{Why does the Neural ISP benefit disproportionately?}

Due to historical mobile computational limits, a traditional ISP does not deconvolve the PSF blur kernel from the sensor capture. It operates locally and treats each step (interpolation, denoising, sharpening) independently. It has no model of the PSF and no ability to invert aberrations. When the PSF is small and well-behaved (large pixels, slow aperture), this does not matter much. When aberrations are severe, the PSF blur remains after the traditional ISP pipeline. Strengthening traditional ISP steps (e.g., using more aggressive chroma filtering) to mitigate PSF blur leads to more severe artifacts and loss of information.

A Neural ISP, by contrast, is trained end-to-end on the specific sensor and optical characteristics and therefore has access to an implicit model of the PSF, CFA pattern, noise statistics, and their interactions. In addition to undoing PSF blur, it learns to invert the combined degradation in a single pass. In the small-pixel era, sensors provide denser sampling, and thus more pixels per unit scene area that the network can exploit during image restoration.

\subsection{Implications for future sensor design}

Our results point toward a design philosophy in which:

\begin{itemize}[leftmargin=*]
    \item Pixel pitch can continue to shrink below 0.5~\um{} with confidence that neural restoration can handle the optical consequences.
    \item Fast apertures are essential at small pixel sizes, not only for light gathering but to avoid the diffraction wall that would otherwise render additional pixels unproductive.
    \item The lens does not need to be perfect. Rather than spending substantial design effort eliminating every last aberration at faster F-numbers, the optical design can be ``good enough'' and rely on the Neural ISP to handle the residual. A wider lens aperture can be used without necessarily requiring a more complex design, taking advantage of the smaller pixels. This expands the design space toward simpler, thinner, or more aggressive optical assemblies.
    \item Co-design of optics and Neural ISP---where the lens is explicitly optimized to produce aberrations that the Neural ISP can efficiently correct---represents a natural next step, and is the subject of ongoing work in our Deep Optics program.
\end{itemize}

The small-pixel era is not solely a challenge; it is also an opportunity. The conventional wisdom that shrinking pixels inevitably degrades image quality assumes a conventional ISP pipeline. With a Neural ISP operating end-to-end from RAW to RGB, smaller pixels provide denser spatial sampling that the network can exploit to achieve higher resolving power, even as the optics become progressively more imperfect.

Across a 2$\times$ range of pixel pitches, the Neural ISP delivers consistent improvements regardless of whether the primary degradation is geometric aberration or diffraction, and at the most challenging 0.35~\um{} configuration these benefits extend to low-SNR scenarios when multi-frame fusion is employed.

For sensor designers, module integrators, and device OEMs, the telephoto physics wall is real, but neural image processing provides a means to climb over it.

\appendix
\section{Supplementary Methods and Source Notes}

\subsection{Wavefront-based PSF simulation}

Since real lens prescriptions from manufacturers are proprietary, we use a Zernike polynomial wavefront fitted to real smartphone camera measurements as a proxy for a telephoto lens assembly. A key aspect of our approach is that we use the same underlying wavefront across all configurations and simply truncate the pupil aperture to match each F-number. As the aperture widens, it captures more of the wavefront's outer regions, where aberrations are strongest. This mirrors what happens physically when a real lens is operated at a faster F-number.

\begin{figure*}[t]
    \centering
    \includegraphics[width=0.95\textwidth]{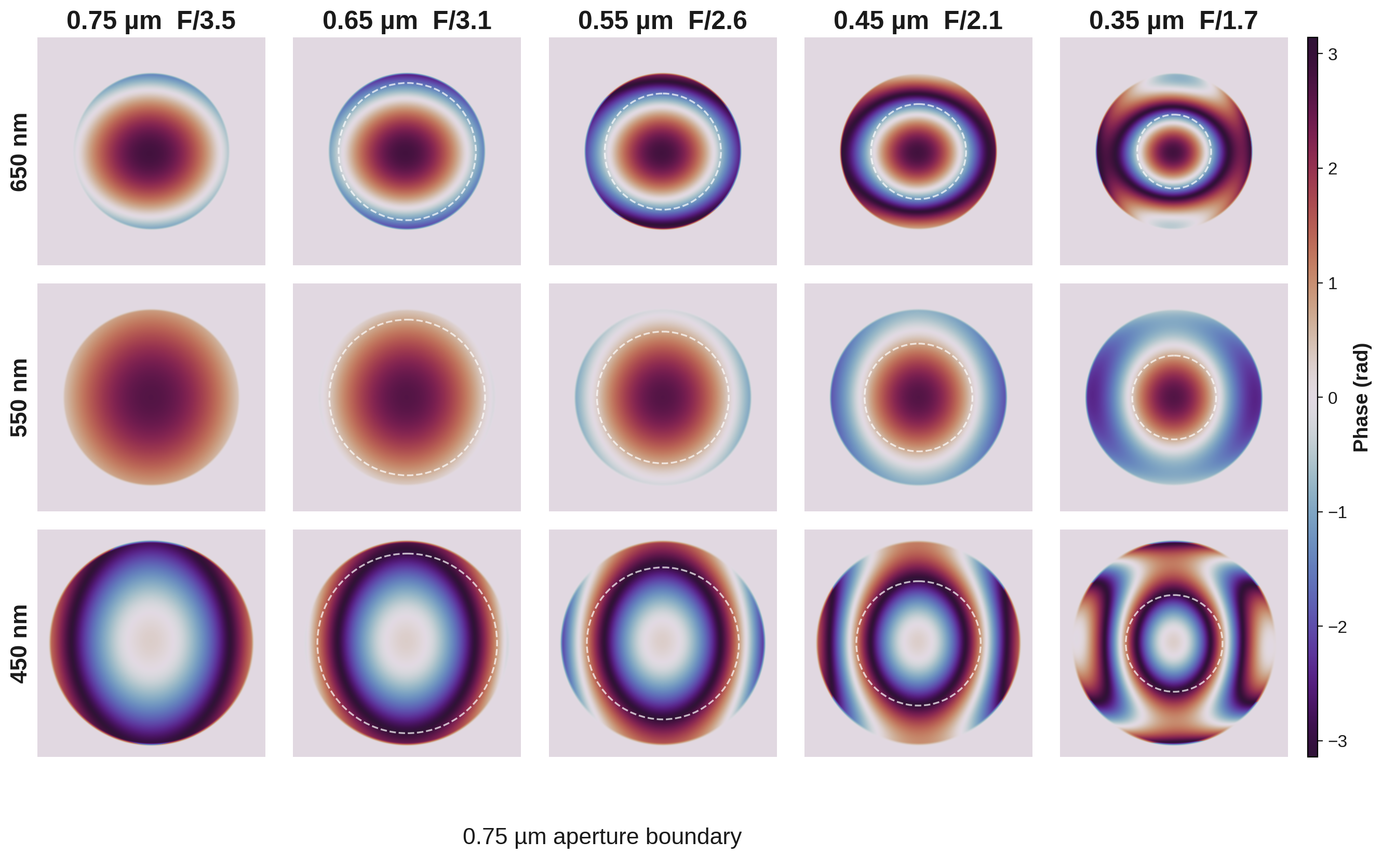}
    \caption{Simulated exit pupil phase for three representative wavelengths at various aperture sizes. The dashed circle indicates a constant effective aperture size for reference; the 0.35~\um, F/1.66 pupil is twice the diameter of the 0.75~\um, F/3.55 pupil.}
    \label{fig:fig12}
\end{figure*}

\begin{figure}[t]
    \centering
    \includegraphics[width=\columnwidth]{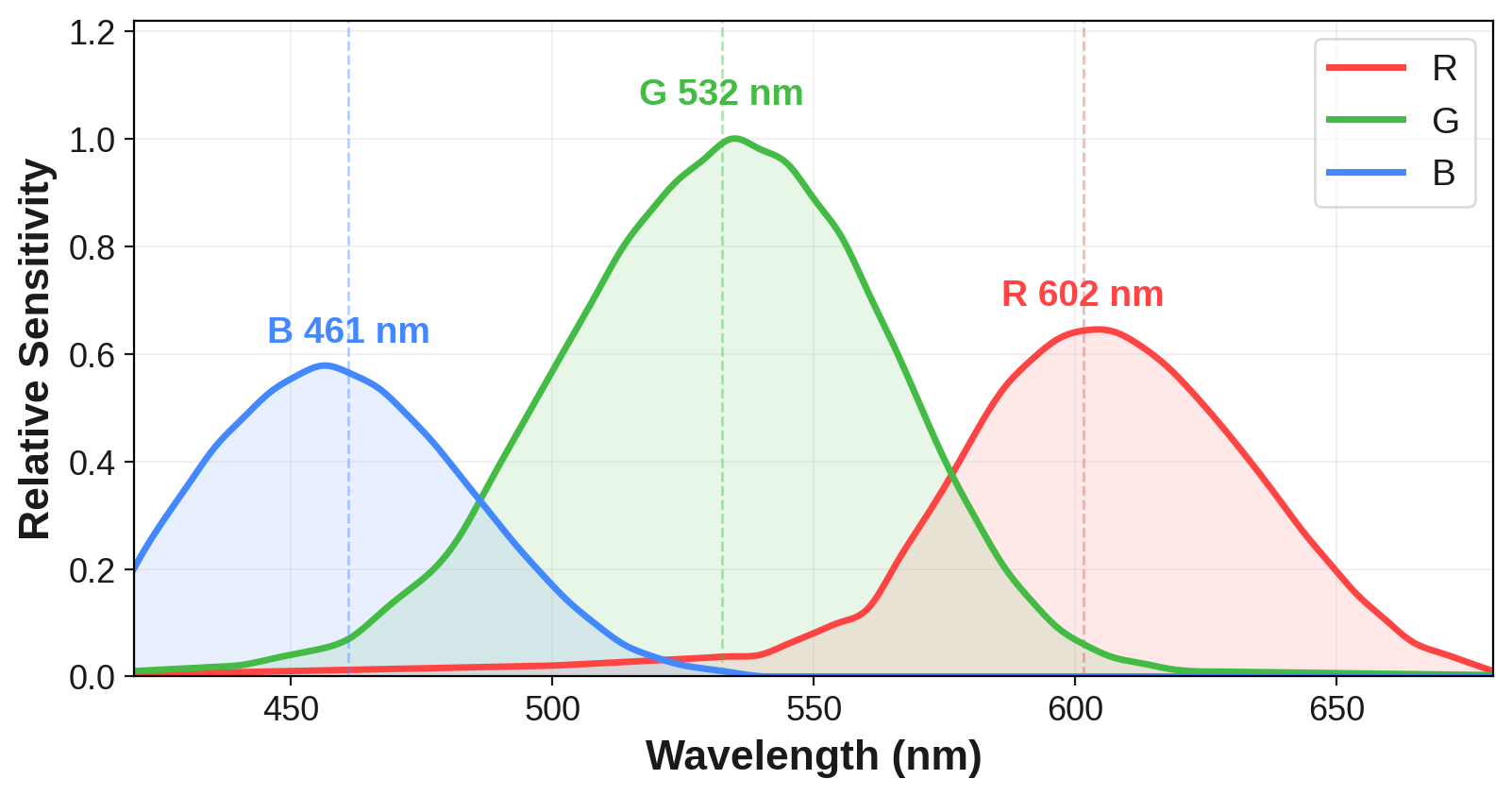}
    \caption{Sensor spectral responsivity used in RGB PSF simulation.}
    \label{fig:fig13}
\end{figure}

We simulated the full exit pupil wavefront across 51 wavelengths spanning 420--680~nm, then computed PSFs using a realistic sensor spectral sensitivity curve (Figure~\ref{fig:fig13}) to generate an incoherent polychromatic PSF. The Zernike coefficients were manually tuned to approximate the aberration profile of a modern folded telephoto module, including controlled levels of spherical aberration, astigmatism, and higher-order terms.

\subsection{SNR normalization, sensor and lens size}

An important property of proportional scaling is that the increase in aperture area exactly compensates for the decrease in pixel area, so photon flux per pixel remains constant across all configurations. Consequently, the signal-to-noise ratio (SNR), which we assume is shot-noise dominated, also remains constant. This means any differences in output quality are not attributable to noise advantages.

With a fixed sensor size, smaller pixels simply increase the megapixel count while maintaining the same field of view. Moving from 0.75~\um{} to 0.35~\um{} more than quadruples the pixel count, and therefore the spatial sampling density of the scene. This also implies that our simulation does not change the focal length or z-height (total track length, or TTL) of the lens, so designs remain physically plausible without modification of the smartphone's industrial design.

A recent trend has been for the TTL to increase in order to support longer focal lengths to improve zoom, or to accommodate more lens elements to improve image quality. This trend is difficult to sustain as camera modules occupy more physical space inside the phone. Although innovations such as periscope lenses and tetraprisms (and larger camera bumps) have enabled longer optical paths, they still consume real estate inside the device. Our simulations show that simply increasing the aperture size on an existing design can fully exploit the benefits of smaller pixels.

\subsection{Modeling of 2$\times$2 OCL microlens effect}

Modern high-resolution sensors use on-chip lenses (OCLs) shared across a 2$\times$2 pixel group. Because each pixel beneath the microlens receives light from a slightly different angular distribution, the chief ray angle (CRA) varies appreciably across the four pixels, ultimately yielding four distinct PSFs within each microlens unit. To model this microlens effect, we introduced small global tilt and coma offsets across the four phase positions, thereby simulating a ``pixel-imbalance effect'' that typically appears as strong edge fringing in mosaic RAW images and remains a well-known challenge in Quad and Hex CFA demosaicing.

\subsection{RAW image synthesis}

After all optical effects, we simulate the sensor. This includes mosaicking through a Hex Bayer CFA pattern and corrupting with sensor noise. We model the noise as additive Gaussian with variance matched to the shot-noise-limited SNR at the operating point (approximately 35~dB, or 15~dB for the low-light evaluation). The CFA pattern transmittance is accounted for via the spectral responsivity in the PSF generation step.

\subsection{Assumptions and simplifications}

We made several simplifying assumptions, including:

\begin{itemize}[leftmargin=*]
    \item \textbf{Pixel-level wave optics:} Diffraction at the pixel aperture and optical crosstalk between adjacent pixels are not modeled, as these effects require detailed knowledge of the sensor's physical structure.
    \item \textbf{Quantum efficiency} is assumed to be preserved across pixel sizes due to advances in sensor technology and the underlying absorptive material remaining the same.
    \item \textbf{The wavefront is not re-optimized per configuration.} A real lens designer would adjust the optical design when targeting a faster F-number. Our approach deliberately holds the aberration profile constant to isolate the effect of aperture scaling, making this a somewhat pessimistic scenario for smaller pixels. This places a higher burden on the Neural ISP for deblurring. (A lens re-optimized per aperture would likely aberrate somewhat less, modestly narrowing the absolute gap.)
\end{itemize}

\subsection{Evaluation Metrics}

We evaluate both the traditional demosaiced output and the Neural ISP restored output using the metrics below. The full-reference metrics (MS-SSIM, LPIPS) are computed against the ground truth, defined as the aberration-free, noise-free RGB rendering of the scene sampled on the target pixel grid.

\begin{itemize}[leftmargin=*]
    \item \textbf{MTF (Modulation Transfer Function) curves:} Measured by feeding simulated sinusoidal grating images at progressively finer spatial frequencies through the full imaging simulation and restoration pipeline, then comparing input-to-output contrast at different frequencies via FFT. This curve quantifies how different spatial frequencies are attenuated through the pipeline.

    \item \textbf{MTF50:} The spatial frequency at which the system MTF drops to 50\%, reported in cycles/mm. Higher values indicate better spatial resolution.

    \item \textbf{MS-SSIM:} Multi-scale structural similarity measures how similar the output image is to the ground truth. Higher is better; averaged over the 159-image test set.

    \item \textbf{LPIPS:} Learned perceptual image patch similarity is another metric for image quality with respect to the ground truth. Lower is better; averaged over the same test set.
\end{itemize}

A critical detail: all metrics are reported in cycles/mm (physical units), not cycles/pixel. This correctly accounts for the higher sampling density of smaller pixels and avoids penalizing them for having more pixels per unit scene area.

\subsection{Holding the Aperture Size Fixed}

\begin{figure*}[t]
    \centering
    \includegraphics[width=0.95\textwidth]{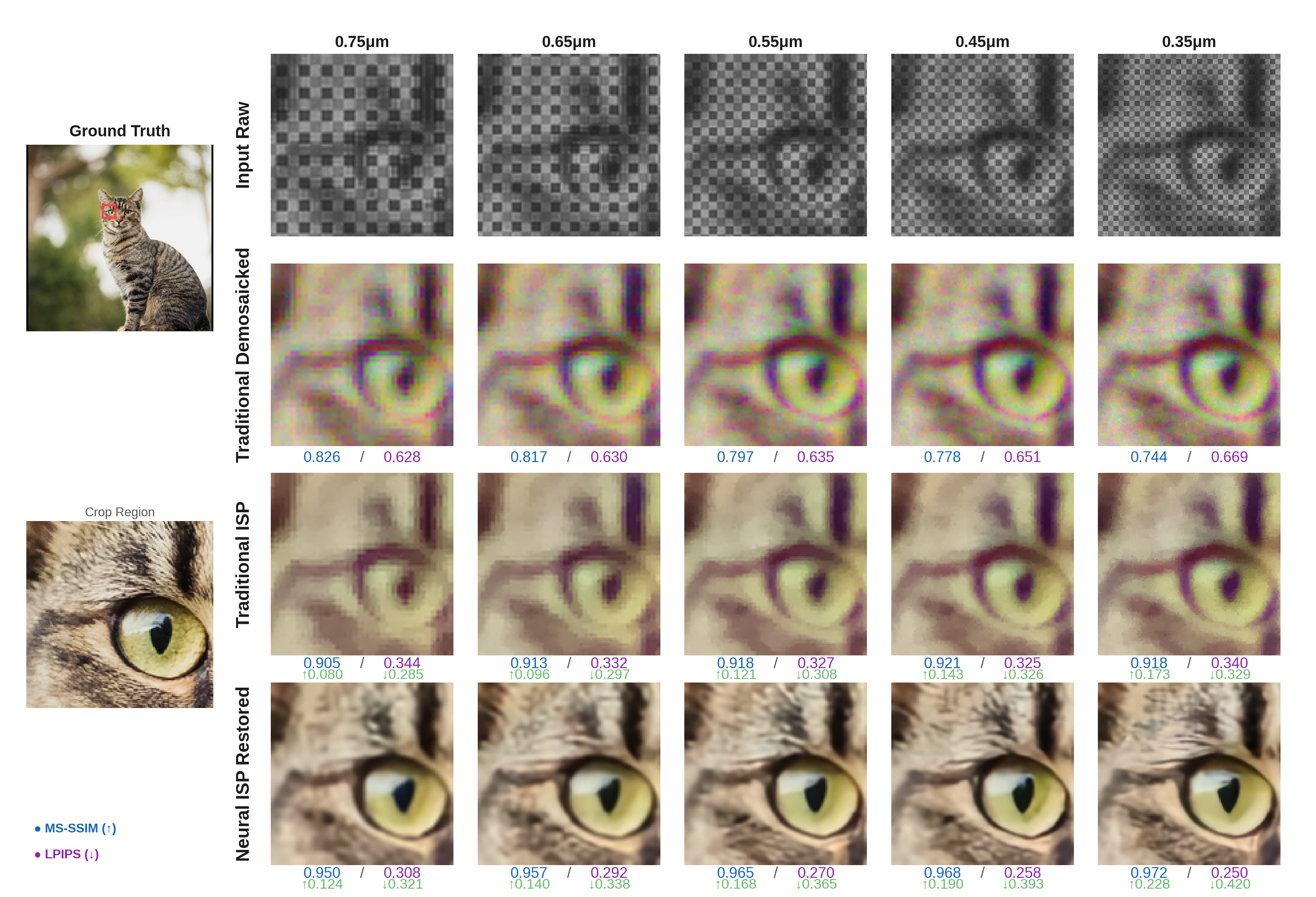}
    \caption{Simulation with decreasing pixel sizes but fixed aperture at F/2.6. The Neural ISP recovers more detail despite the diffraction-limited PSF in the smaller-pixel scenarios.}
    \label{fig:fig14}
\end{figure*}

We also tested the case where the pixel size decreased but the aperture size stayed constant at F/2.6. The results are shown in Figure~\ref{fig:fig14} and Table~\ref{tab:caseB}.

\begin{table*}[t]
\centering
\small
\caption{Fixed-aperture ablation (Case B): perceptual metrics at constant F/2.6 as pixel pitch decreases (single-frame, 35~dB). MS-SSIM: higher is better; LPIPS: lower is better. As the diffraction spot covers more pixels at smaller pitch, the demosaiced output degrades (and the traditional ISP LPIPS worsens at the smallest pitch), whereas the Neural ISP continues to improve---demonstrating robustness whether the dominant degradation is geometric aberration or diffraction.}
\label{tab:caseB}
\begin{tabular}{c rrr rrr}
\toprule
& \multicolumn{3}{c}{MS-SSIM $\uparrow$} & \multicolumn{3}{c}{LPIPS $\downarrow$} \\
\cmidrule(lr){2-4}\cmidrule(lr){5-7}
Pitch (\um) & Demosaiced & Traditional & Neural & Demosaiced & Traditional & Neural \\
\midrule
0.75 & 0.791 & 0.857 & 0.955 & 0.533 & 0.380 & 0.249 \\
0.65 & 0.790 & 0.870 & 0.963 & 0.529 & 0.356 & 0.222 \\
0.55 & 0.783 & 0.880 & 0.961 & 0.532 & 0.347 & 0.202 \\
0.45 & 0.774 & 0.886 & 0.967 & 0.547 & 0.351 & 0.188 \\
0.35 & 0.762 & 0.885 & 0.973 & 0.558 & 0.374 & 0.168 \\
\bottomrule
\end{tabular}
\end{table*}

\subsection{Telephoto Camera Specifications from GSMArena}

Table~\ref{tab:gsmarena} lists the telephoto camera specifications used as data points in Figure~\ref{fig:fig1}. All entries are sourced from GSMArena (\url{https://www.gsmarena.com}); the device-specific URL is constructed by appending the slug in the final column to the base URL with the prefix \texttt{/} and the \texttt{.php} suffix.

\begin{table*}[t]
\centering
\small
\caption{Telephoto camera specifications from GSMArena. Full URL: \texttt{https://www.gsmarena.com/<slug>.php}.}
\label{tab:gsmarena}
\resizebox{\textwidth}{!}{%
\begin{tabular}{lccccccl}
\toprule
Device & MP & F/\# & Focal Length & Sensor Size & Pixel Pitch & Optical Zoom & Slug \\
\midrule
iPhone 7 Plus      & 12  & 2.8 & 56~mm  & 1/3.6"  & 1.0~\um& 2$\times$    & \texttt{apple\_iphone\_7\_plus-8065} \\
Galaxy Note 8      & 12  & 2.4 & 52~mm  & 1/3.6"  & 1.0~\um   & 2$\times$    & \texttt{samsung\_galaxy\_note8-8505} \\
Huawei P30 Pro     & 8   & 3.4 & 125~mm & 1/4.0"  & 1.22~\um& 5$\times$    & \texttt{huawei\_p30\_pro-9635} \\
iPhone 11 Pro      & 12  & 2.0 & 52~mm  & 1/3.4"  & 1.0~\um   & 2$\times$    & \texttt{apple\_iphone\_11\_pro-9847} \\
Galaxy S20 Ultra   & 48  & 3.5 & 103~mm & 1/2.0"  & 0.8~\um   & 4$\times$    & \texttt{samsung\_galaxy\_s20\_ultra-10084} \\
Pixel 6 Pro        & 48  & 3.5 & 104~mm & 1/2.0"  & 0.8~\um   & 4$\times$    & \texttt{google\_pixel\_6\_pro-10918} \\
Honor Magic4 Pro   & 64  & 3.5 & 90~mm  & 1/2.0"  & 0.7~\um   & 3.5$\times$  & \texttt{honor\_magic4\_pro-11390} \\
Pixel 8 Pro        & 48  & 2.8 & 113~mm & 1/2.55" & 0.7~\um   & 5$\times$    & \texttt{google\_pixel\_8\_pro-12545} \\
Xiaomi 14 Ultra    & 50  & 2.5 & 120~mm & 1/2.51" & 0.7~\um   & 5$\times$    & \texttt{xiaomi\_14\_ultra-12827} \\
Xiaomi 15 Ultra    & 200 & 2.6 & 100~mm & 1/1.4"  & 0.56~\um  & 4.3$\times$  & \texttt{xiaomi\_15\_ultra-13657} \\
Honor Magic8 Pro   & 200 & 2.6 & 85~mm  & 1/1.4"  & 0.56~\um  & 3.7$\times$  & \texttt{honor\_magic8\_pro\_5g-14231} \\
Oppo Find X9 Pro   & 200 & 2.1 & 70~mm  & 1/1.56" & 0.5~\um   & 3$\times$    & \texttt{oppo\_find\_x9\_pro-14094} \\
\bottomrule
\end{tabular}%
}
\end{table*}

\end{document}